\documentclass[10pt,leqno]{article}
\usepackage[utf8]{inputenc}
\usepackage{graphicx}
\usepackage{amstext,amsmath,amssymb,amsfonts,bbm,slashed}
\usepackage{amsthm}
\usepackage{xcolor}
\usepackage[left=2cm,right=2cm,top=2cm,bottom=2cm]{geometry}
\usepackage{tikz}
\usepackage{epigraph}
\usepackage{cite}
\usepackage{mdframed}
\usepackage{authblk}

\newtheorem{definition}{Definition}
\newtheorem{theorem}{Theorem}

\newtheorem{proposition}{Proposition}

\newcommand{\tr}{\operatorname{Tr}}

\newcommand{\be}{\begin{equation}}
\newcommand{\ee}{\end{equation}}
\newcommand{\bea}{\begin{eqnarray}}
\newcommand{\eea}{\end{eqnarray}}

\newcommand{\iPop}{{}^iP}

\title{Tensor Models: extending the matrix models structures and methods}
\author{ St\'ephane Dartois\footnote{stephane.dartois@outlook.com}}
\affil{LPTM, UCP - CNRS, Université Cergy-Pontoise,
2 avenue A. Chauvin, Pontoise
95302 Cergy-Pontoise, France }

\begin{document}
\maketitle

\vspace{20pt}

\begin{abstract}
In this text we review a few structural properties of matrix models that should at least partly generalize to random tensor models. We review some aspects of the loop equations for matrix models and their algebraic counterpart for tensor models. Despite the generic title of this review,  we, in particular, invoke the Topological Recursion. We explain its appearance in matrix models. Then we state that a family of tensor models provides a natural example which satisfies a version of the most general form of the topological recursion, named the \emph{blobbed} topological recursion. We discuss the difficulties of extending the technical solutions existing for matrix models to tensor models. Some proofs are not published yet but will be given in a coming paper, the rest of the results are well known in the literature      
\end{abstract}

\vspace{10pt}
\section{Introduction}

Since Razvan Gurau and collaborators have discovered \emph{colored} random tensor models and their $1/N$ expansion imitating the $1/N$ expansion of random matrix models, the field of random tensor models has been growing fastly. Indeed the introduction of colors cured several problems which prevented the early tensor models introduced in \cite{Ambjorn} to grow further as an approach to higher dimensional quantum gravity.  \\
From a mathematical point of view, these colored tensor models are generating functions of a special class of piecewise linear ($PL$) manifolds in dimension $d$ called \emph{GEM} \cite{FeGa}. 
These $PL$-manifolds are counted with respect to their degree (which, from a combinatorial standpoint, generalizes the genus) and their discrete volume.\\
The intense scientific activity of the field leads to new questions for mathematical physicists, most of them being rather challenging. A good amount of these new results and problems come from trying to extend what we know about two dimensional tensor models (= matrix models) to higher dimensional models. There exists a lot of examples supporting such a claim. One thinks about the extension of the $1/N$ expansion (see for instance \cite{complete1/N}), the computation of first and second order \cite{bijection,KOR}, the extension of double scaling limits - for instance in \cite{DaGR,BGRT} -, saddle point techniques \cite{NDE} and the study of Schwinger-Dyson equations by combinatorial means \cite{uncolored, BGRT}. Questions about the possible integrable properties arose recently \cite{Dartois}. If most tensor models are probably not integrable, it is likely that the rich structure of integrable matrix models are generalized in the context of tensor models (a first example is provided by \cite{Dartois}). \\
Among all these results, one misses the extension of one of the most elegant and powerful discovery in matrix models: the topological recursion. The topological recursion has been settled for the study of loop equations of the Hermitian one matrix model\cite{Eyn04}. It since has been linked to the algebraic geometry of surfaces \cite{PhDOr}. Then the topological recursion has been used to gain new insights in topological string theory, invariants of moduli spaces, combinatorics, integrable systems and knot theory \cite{EO08,EynLect,Eyn}.\\

\bigskip       

In this review paper we focus on this idea. In a first part we recall how the topological recursion arises in the study of the loop equations of a Hermitian one matrix model. We show how this is linked to the so-called Virasoro constraints. In a second part, we describe the Schwinger-Dyson equations of a generic tensor model. We explain why an extension of the formalism of the loop equations, if not excluded, is not obvious. Since we cannot make progress in the generic case we show in a third and fourth part what can be done in a simpler case called the Quartic Melonic Tensor Model. We indeed show that this particular model satisfy a version of the Blobbed Topological Recursion in a specific series of dimensions $d=4k+2$. This induces that all the structures attached to the blobbed version of the topological recursion can be found in this particular model.   Nevertheless, there still are difficulties left to treat to conclude that we solved this model.          
The conclusion is then devoted to the presentation of a program whose aim is the discovery of Topological Recursion like structures in tensor models. 

\section{Loop equations for matrix models and how to solve them.}\label{sec:LEMM}
\subsection{Loop equations}
Matrix models obey a set of constraints that appear in the literature under many different names\cite{Eyn,DfGiZJ,EynLect}. One names them Loop equations, Virasoro constraints, Schwinger-Dyson equations, Cut-and-Join equations, Tutte equations and so on... This is due to the fact that the structure of this set of equations appears not only in matrix models but in a wide variety of domains of mathematics, physics and combinatorics\cite{Eyn}. Although this topics is fascinating, it is above all incredibly rich. As a consequence it is not possible to make a complete presentation of the subject here and this for two reasons. The first (and main) is that the author is not competent enough. The second one is that it would fall way beyond the scope of this review paper. So, in this text we make the choice of presenting only the small part of it that is the closest in spirit to what one would extend to tensor models. \\

Loop equations\footnote{from now on we use this name for the constraints satisfied by matrix models.} are nothing more than the consequences of the invariance of a Lebesgue measure under translations. This particular feature turns out to be important since this implies that these equations are true even non-perturbatively. However, when it comes to finding solutions (exact or not), one often restricts to perturbative solutions. \\ 
These sets of equations are obtained as follows\cite{Ambjorn90,EO08}. Consider, for instance, a Hermitian matrix model, whose partition function writes
\begin{equation}
Z_{1MM}[\{t_p\}]=\int_{H_N}dM \exp\left( -N\tr V(M)\right).
\end{equation}
From the choice of $V(M)$ depends the possible interpretations of this integral. Indeed for a generic $V(x)=\sum_{p\ge 1} t_p x^p$ it is not obvious to state the needed conditions on the $t_p$'s for the integral to converge. However, as long as one understands this integral perturbatively, as an expansion around all $t_{p\neq 2}=0$, $t_2= 1/2$, one can forget about these conditions\footnote{but if these conditions are satisfied, the derivation presented here is still valid.}.
The integral presented above is invariant under change of variables. In particular, this is invariant under the change of variables $M \rightarrow M'= M+\epsilon (x-M)^{-1}$. The Jacobian $J$ is, up to the first order in $\epsilon$, $J\simeq \sum_{p,a,q,b}\delta_{pa}\delta_{qb}+\epsilon \partial_{M_{pq}}(x-M)^{-1}_{ab}$. The exponential term is at first order $\exp\left( -N\tr V(M)\right)(1-N\epsilon\tr(V'(M)(x-M)^{-1}))$. 
Then the invariance implies
\begin{equation}\label{eq:1pt}
\left\langle \tr(x-M)^{-1}\tr(x-M)^{-1}\right\rangle-\left\langle\tr \frac{NV'(M)}{(x-M)}\right\rangle =0.
\end{equation}
This is the first of a series of equations that are obtained by making the more general changes $M'= M+\epsilon (x_1-M)^{-1}\tr(x_2-M)^{-1}\ldots \tr(x_n-M)^{-1}$. Denoting 
\begin{align}
&\bar{W}_n(x_1,\ldots, x_n )=\left\langle \tr(x_1-M)^{-1}\tr(x_2-M)^{-1}\ldots \tr(x_n-M)^{-1} \right\rangle.\\
&W_n(x_1,\ldots, x_n )=\left\langle \tr(x_1-M)^{-1}\tr(x_2-M)^{-1}\ldots \tr(x_n-M)^{-1} \right\rangle_c.
\end{align}
Using these objects, equation \eqref{eq:1pt} writes,
\begin{equation}
W_{2}(x,x)+W_1(x)^2-\left\langle \tr\frac{NV'(M)}{(x-M)}\right\rangle=0,
\end{equation}
while we derive from the more general changes of variables,
\begin{align}
&W_{n+1}(x_1,x_1,x_2,x_3,\ldots,x_n)+ \\ &\sum_{i\ge 2}\partial_{x_i}\frac{W_n(x_1,\{x_p\}_{p\neq 1,i})-W_n(\{x_p\}_{p\neq 1})}{x_1-x_i}
-\left\langle \tr\frac{NV'(M)}{(x_1-M)}\tr(x_2-M)^{-1}\ldots \tr(x_n-M)^{-1}\right\rangle=0.
\end{align} 
When considering perturbative solutions, one can expand the $W$'s in term of $N$ the size of the matrix\footnote{this is not always possible for convergent matrix integrals that can contain term of the form $e^{-A/N^2}$}. The prototypical type of expansion is 
\begin{equation}
W_n(x_1,\ldots, x_n)=\sum_{h\ge 0}N^{2-2g-n}W_n^g(x_1,\ldots, x_n).
\end{equation}
One then gets a series of equations indexed by $(g,n)$. For instance,
\begin{align}
&W_{n+1}^{g-1}(x_1,x_1,x_2,x_3,\ldots,x_n)+ \\ &\sum_{i\ge 2}\partial_{x_i}\frac{W_n^g(x_1,\{x_p\}_{p\neq 1,i})-W_n^g(\{x_p\}_{p\neq 1})}{x_1-x_i}
-\left\langle \tr\frac{NV'(M)}{(x_1-M)}\tr(x_2-M)^{-1}\ldots \tr(x_n-M)^{-1}\right\rangle^g_{n-1}=0
\end{align}
Bertrand Eynard, Nicolas Orantin and collaborators \cite{Eyn04,PhDOr} constructed a formula to compute the $W$'s. This is now understood as relying on the fact that the equation on $W_1^0$ defines an affine plane curve\cite{PhDOr, Eyn}. In fact, if one considers $W_1^0$ and $x$ as two complex indeterminates, the first equation is (in the simplest case) a polynomial in these two indeterminates in $\mathbb{C}^2$. This polynomial generically vanishes on a surface in $\mathbb{C}^2$. Up to subtleties, the choice of variables and polynomial made to describe this surface correspond to a choice of complex structure on it\cite{Mir}. Finding the expression $W_1^0(x)$ corresponds to writing a description of this surface as a sheeted surface. 

\subsection{Topological Recursion in the $1$-cut case}
In the next part we restrict to the so called $1$-cut case\footnote{We also make an ``under the carpet'' assumption: the potential is polynomial.}. So to say $x$ lives in the complex plane and $W_1^0(x)$ has one cut somewhere in this complex plane so that it is naturally defined on two copies of the complex plane\cite{Eyn04, Eyn}. The corresponding surface is the Riemann surface of $W_1^0$ and is the sphere. Using uniformizing coordinates, it is possible to write $W_1^0$ as a function on the sphere seen as $\mathbb{C}\cup \{\infty\}$.
These coordinates are known in physics under the name of Joukowsky coordinates. They write
\begin{align}
& x(z)=\frac{a+b}{2}+\frac{a-b}{4}(z+1/z) \\
& z(x)=\frac{2}{a-b}\left(x - \frac{a+b}{2} +\sqrt{(x-a)(x-b)}\right),
\end{align}   
where $a,b$ are the endpoints of the cut. A choice of determination for the square root is a choice of sheet for $x$ where to live, and so is a choice of one of the two standard charts for the sphere $\mathbb{C}\cup \{\infty\}$. Through this transformation, the 
cut is mapped to the unit circle with $a$ and $b$ being sent to $-1$ and $+1$ and the transition from one chart to the other is given by the transition map $\iota(z)=1/z$. 
\\
In order to study properly the analytic properties of the solutions one needs\cite{Schli,EO08} to define the forms 
\begin{equation}
\omega_{n}^g(z_1,\ldots, z_n):= W_n^g(x_1(z_1),\ldots, x_n(z_n))dx_1\otimes \ldots \otimes dx_n. 
\end{equation} 
by pull-backing the $W$'s by the function $x(z)$ and turning these pull-backs into sections of tensor products\footnote{We again put some symmetry assumptions under the carpet, it is not just a tensor product \cite{Eyn}.} of the canonical bundle. The equations on the $W$'s directly translate into equations on the $\omega$'s. In these coordinates, these equations are equivalent to a set of linear equations and quadratic equations. The first are called linear loop equations\cite{Eyn04, BEO} and state for $2g-2+n>0$ that
\begin{equation}
S\omega_n^g(z_1,\ldots, z_n):=\omega_{n}^g(z_1,\ldots, z_n)+\omega_{n}^g(\iota(z_1),\ldots, z_n)=0.
\end{equation}     
While the second are called quadratic loop equations and state that the quantity often denoted $\mathcal{Q}^g_n(z_1;z_2,\ldots, z_n)$ and defined by
\begin{equation}
\mathcal{Q}^g_n(z_1;z_2,\ldots, z_n):=\omega_{n+1}^{g-1}(z_1,\iota(z_1),z_2,\ldots, z_n) +\sum_{\substack{J\subseteq I \\ h+h'=g}}\omega_{|J|+1}^h(z_1,J)\otimes \omega_{|I-J|+1}^{h'}(\iota(z_1),I-J),
\end{equation}
where $I=\{z_2,\ldots,z_n\}$, has a double zero at $z_1=\pm 1$. When satisfied, these conditions allow one to compute any $\omega_n^g$ with a recursive residue formula
\begin{align}\label{eq:mainformula1}
&\omega_n^g(z_1,\ldots,z_n)=\sum_{\pm 1}\underset{z\rightarrow \pm 1}{\textrm{Res}}K(z,z_1)\bigl\{\tilde{\mathcal{Q}}^g_n(z_1;z_2,\ldots, z_n) \bigr\} \\
&:=\sum_{\pm 1}\underset{z\rightarrow \pm 1}{\textrm{Res}}K(z,z_1)\bigl\{ \omega_{n+1}^{g-1}(z_1,\iota(z_1),z_2,\ldots, z_n) +\sum_{\substack{J\subseteq I \\ h+h'=g}}^{'}\omega_{|J|+1}^h(z_1,J)\otimes \omega_{|I-J|+1}^{h'}(\iota(z_1),I-J \bigr\},
\end{align}
where $\sum^{'}$ indicates that the sum only involves terms without $\omega_{1}^0$. $K(z,z_1)$ is a kernel that is not symmetric in its variables. This kernel is constructed out of $\omega_{1}^0$ and $\omega_{2}^0$\cite{EynLect}. These forms are initial conditions for the recursion. One consequently has to determine the Riemann surface (with its complex structure) described by the polynomial in $(W_1^0, x)$, but also a bi-differential on it given by $\omega_2^0(z_1,z_2)$ and that is computed from the loop equations.
This bi-differential has the property of having a double pole at $z_1=z_2$. This why in some sense it allows to write a Cauchy formula\cite{EynLect} that leads after some work to equation \eqref{eq:mainformula1}. This formula is at the core of the formalism called the Topological Recursion. If we have restricted here to the matrix models point of view\footnote{In fact, even further, we restricted to a small subset of the matrix models point of views.}, the reader is warned that this structure is way more general, and forms $\omega_n^g$ can be associated to any suitable choices of $\omega_1^0$ and $\omega_2^0$ coming from a physical, combinatorial or mathematical problem. \\

\bigskip

Let us now come back to the $W$'s, that up to some change of variables, we are now able to compute.
As we defined them they write as some correlators of a matrix model,
\begin{equation}
W_n(x_1,\ldots, x_n )=\left\langle \tr(x_1-M)^{-1}\tr(x_2-M)^{-1}\ldots \tr(x_n-M)^{-1} \right\rangle_c.
\end{equation} 
If we focus on their expansion at $x_i=\infty$ we end with,
\begin{equation}
W_n(x_1,\ldots, x_n )=\sum_{\{p_i\}_{i=1}^n}\frac{\left\langle \tr(M^{p_1})\tr(M^{p_2}\ldots \tr(M^{p_n}))\right\rangle_c}{x_1^{p_1+1}x_2^{p_2+1}\ldots x_n^{p_n+1}}.
\end{equation} 
Then in the perturbative setting, the coefficient of $x_1^{-p_1-1}x_2^{-p_2-1}\ldots x_n^{-p_n-1}$ is actually counting the number of maps with $n$ marked boundary of length $p_1, p_2, \ldots, p_n$. More geometrically, it counts the number of $PL$ cobordisms between a collection of $n$ $PL$ $\mathbb{S}^1$. The $W$'s are then generating functions for these numbers. Constructing these generating functions of observables of the matrix models is easily possible because the observables are indexed by a collection of integers (the length of the $PL$ $\mathbb{S}^1$).  

\subsection{Virasoro constraints in matrix models}
The first equation \eqref{eq:1pt} actually contains the constraints often called Virasoro constraints. Indeed, if we perform the expansion at $x_1=\infty$ as above we obtain 
\begin{align}
&\langle \tr(x-M)^{-1}\tr(x-M)^{-1}\rangle-\langle \tr\frac{NV'(M)}{(x-M)}\rangle =
\\&\sum_p x^{-p-1}\bigl(\sum_{k=0}^{p-1} \langle \tr(M^k)\tr(M^{p-1-k})\rangle -N\langle \tr (M^p V'(M)) \rangle\bigr)\nonumber.
\end{align}
Introducing operators, 
\begin{align}
L_p=\frac{1}{N^2}\sum_{k=0}^{p-1}\frac{\partial^2}{\partial t_k \partial t_{p-1-k}}- \sum_{k\ge 0}kt_k\frac{\partial}{\partial t_{k-1+p}},
\end{align} 
we have that the equation \eqref{eq:1pt} rewrites as
\begin{equation}
\hat{W}(x)Z[\{t_p\}]=0,
\end{equation}
where $\hat{W}(x)$ is the operator (formally) defined as 
\begin{equation}
\hat{W}(x)=\sum_{p\ge 0}\frac{1}{x^{p+1}}L_p.
\end{equation}
Since the $L_p$'s form a Virasoro algebra, the loop equations are also often called Virasoro constraints. The operator $\hat{W}$ can be extended to the whole Riemann sphere using the $z$ coordinates.
\section{Constraint equations for tensor models} 
\subsection{Schwinger-Dyson equations of tensor models}
In this subsection we consider a generic tensor model of the form
\begin{equation}
Z[N,\{t_{\mathcal{B}}\}]= \int dT d\bar{T} \exp\Bigl(-N^{d-1} \sum_{\mathcal{B}}\frac{N^{-\frac{2}{(d-2)!}\omega(\mathcal{B})}}{|Aut(\mathcal{B})|} t_{\mathcal{B}}\mathcal{B}(T, \bar{T})\Bigr)
\end{equation}
where the $t_{\mathcal{B}}$'s are coupling constants associated to the tensor invariant indexed by the $(d-1)$ colored graphs $\mathcal{B}$\cite{uncolored}.
One can introduce the formal generating series of observables
\begin{equation}
\mathfrak{G}[\{q_{B_{\bullet}}\}]=\sum_{B_{\bullet}}q_{B_{\bullet}}B_{\bullet}(T,\bar{T}).
\end{equation}
$B_{\bullet}$ is a graph with a marked white vertex, this means that it represents a linear form over the vector space of the $\bar{T}$'s. When evaluated at $\bar{T}$ this form gives the corresponding invariant $B$.
We now make the change of variables $T\rightarrow T'= T+\epsilon \mathfrak{G}[\{q_{B_{\bullet}}\}]$, inducing the change of variables $\bar{T}\rightarrow \bar{T}'=\bar{T}+\bar{\epsilon}\overline{\mathfrak{G}[\{q_{B_{\bullet}}\}]}$. Computing the Jacobian of the transformation at first order in $\epsilon$ we obtain that it writes 
\begin{equation}
|J|=1+\epsilon\tr_{T\oplus \bar{T}}\left(\sum_{B_{\bullet}}q_{B_{\bullet}} \partial_{T_{a_1\ldots a_d}}B_{\bullet}(T,\bar{T})_{a_1\ldots a_d}\oplus e^{-2i\theta} c.c.\right) + O(\epsilon^2)).
\end{equation} 
We used in particular $\epsilon = |\epsilon|e^{i\theta}$. As explained in \cite{Guconstraints, DarPhD} the derivatives terms appearing in the trace can all be represented graphically. After taking the trace, they become invariants obtained by contracting the lines between the marked vertex and the vertex representing the $T$ (resp. $\bar{T}$) the derivative operator acted on. 
This change of variables also induces a change in the potential of the integrand, that writes, up to first order in epsilon, 
\begin{equation}
V(T,\bar{T})+\epsilon \sum_{a_1\ldots a_d}\partial_{T_{a_1\ldots a_d}}V(T,\bar{T})\sum q_{B_{\bullet}}B_{\bullet}(T,\bar{T})_{a_1\ldots a_d}+ c.c. .
\end{equation} 
Each term proportional to $q_{B_{\bullet}}$ represents an invariant obtained by contracting the marked vertex of $B_{\bullet}$ and the vertex of a $\mathcal{B}(T,\bar{T})$ in the potential $V(T,\bar{T})$ acted on with the derivative operator. This leads to a full series of terms denoted $\mathcal{B}\star_v B_{\bullet}$ proportional to $q_{B_{\bullet}}$. This can be expressed using differential operators\cite{Guconstraints} of the $t_{\mathcal{B}}$, $L_{B_{\bullet}}$ associated\footnote{In the sketch of derivation presented here one ends with $L_{B_{\bullet}}$ that are direct sums of the differential operators presented in \cite{Guconstraints}. } to each $q_{B_{\bullet}}$. We have from the fact that the integral does not change under change of variables that,
\begin{equation}
\hat{\mathfrak{G}}[\{q_{B_{\bullet}}\}]Z[N,\{t_{\mathcal{B}}\}]:=\left(\sum_{B_{\bullet}}q_{B_{\bullet}}L_{B_{\bullet}}\right)Z[N,\{t_{\mathcal{B}}\}]=0.
\end{equation}    
This is very similar to what is obtained in the context of matrix models. However we cannot keep track of the full structure of the observable $B$ with a single power of a complex variable anymore. 
That is the main obstacle to an extension of the Topological Recursion to general tensor models.
Certainly one can try to forget about some structure of the observables to keep only what can be encoded with one single integer. For instance one can think of the volume of the observable (its number of vertices), or the length of one of the face going through the marked vertex. In the three dimensional case, the last choice would provide generating functions of proper normal embedding of a surface in a three dimensional $PL$-manifold; while the first leads to generating functions of $PL$-manifolds with respect to the volume of their boundaries. Doing so a lot of $q_{B_{\bullet}}$ collapse to the same value, and we can, \textit{a priori}, obtain functions defined on the complex numbers. Unfortunately it is not obvious to get the corresponding properties of these generating functions even in these simplified cases. 

\bigskip

So in order to investigate the structure of tensor models and their Schwinger-Dyson equations a bit more while slowly moving towards a generalization of the Topological Recursion to higher dimensional cases, we review, in the next section, a simpler model.  
 
\section{The Quartic Melonic model}
\subsection{Introduction of the model and Loop equations}
We consider the (almost) simplest interacting tensor model in dimension $d$. Its potential contains a quadratic term and all the quartic melonic interaction terms, symmetrically weighted. In this review it is named Quartic Melonic Tensor Model (QMTM). Its partition function writes\cite{Beyond,NDE},
\begin{equation}
\mathcal{Z}[N,\lambda] = \int_{(\mathbb{C}^{N})^{\otimes D}} dT d\bar{T} \exp\Biggl(-N^{D-1}\biggl(\frac{1}{2}(\bar{T}\cdot T) +\frac{\lambda}{4}\sum_c 
(\bar{T}\cdot_{\hat{c}} T)\cdot_c (\bar{T}\cdot_{\hat{c}} T)\biggr)\Biggr)
\end{equation}  
After resummation of the first order and thanks to a Hubbard-Stratanovitch transformation, it is possible to rewrite the partition function, up to a proportionality factor as \cite{NDE, DarPhD},
\begin{equation}
Z[\alpha, N]=\int_{f, H_N^d} \prod_{c=1}^d dM_c e^{-\frac{N}{2}\sum_{c=1}^d \tr(M_c^2)}e^{-\tr\log_2\Bigl[\mathbbm{1}^{\otimes d}-\frac{\alpha^p}{N^{\frac{d-2}{2}}}\sum_{c=1}^d\mathcal{M}_c\Bigr]},
\end{equation}
where the $M_c$ are a family of $d$ matrices, while the $\mathcal{M}_c$ are tensor products of the form 
\begin{equation}
\mathcal{M}_c=\mathbbm{1}^{\otimes (c-1)}\otimes M_c \otimes \mathbbm{1}^{\otimes (d-c)}.
\end{equation}
$\log_2$ means that we consider the formal series in $\alpha$ associated to the $\log$ and that we forget the first term of this series. $\alpha$ is the resummed coupling constant whose value can be expressed from $\lambda$\cite{NDE}.
This resummed model, is a multi-matrix model that satisfies loop equations. These equations can be derived in a similar fashion than in section \ref{sec:LEMM}. However one should point out a first problem in this  tentative to generalize the topological recursion. The potential here contains non-trivial $N$ terms spoiling the topological expansion of observables of these models\footnote{which is needed to have the necessary analytic properties of the $W$'s.}. Fortunately, in a series of dimensions $d=4k+2$, $k\in \mathbb{N}$ \cite{DarPhD}, these $N$ terms do not destroy the topological expansion as long as we consider that the interaction terms now represent cells with non trivial topology (\textit{i.e.} not just discs as in usual matrix models, but any surface with at least one boundary). This idea has been investigated in a work of \cite{BoStuffed} in the context of generalized Hermitian one matrix model. Coming back to the loop equations, if we focus for instance on the $d=6$ dimensional case, even though they are more complicated and need the introduction of some multi-dimensional integral terms, it is still possible to write them. These additional multi-dimensional integral terms are due to the fact that the potential writes in term of tensor products of the $M_c$ and not in term of the $M_c$ alone. These loop equations write 
\begin{align}\label{eq:LEgeneral}
&\sum_{\substack{g\ge h\ge 0\\ \mathbf{q}+\mathbf{r}=\mathbf{k} | \mathbf{q},\mathbf{r},\mathbf{k} \in \mathbb{N}^{d=6}}} W_{e_1+\mathbf{q}}^h(x,x_{\mathbf{q}})W_{e_1+\mathbf{r}}^{g-h}(x,x_{\mathbf{r}})+W_{2e_1+\mathbf{k}}^{g-1}(x,x, x_{\mathbf{k}}) + \sum_{i\in [\![1,k_1]\!]} \partial_{x^{(1)}_i}\frac{W^g_{e_1+\mathbf{k}}(x,x_{\mathbf{k}}\setminus {x^{(1)}_i})-W^g_{e_1+\mathbf{k}}(x_{\mathbf{k}})}{x-x_i} \nonumber \\
&-\oint\frac{d\zeta_1}{2i\pi}\frac{\zeta_1 W_{e_1+\mathbf{k}}^g(\zeta_1, x_{\mathbf{k}})}{x-\zeta_1} \nonumber \\
&+\sum_{p\ge 2}\frac{\alpha^p}{p}\sum_{\vec{q}\in F_p}\binom{p}{\vec q}^{\backsim}\sum_{\substack{J\vdash \mathcal{C} \\ \{h_i\ge 0\} \\ g=p -|J|+4+\sum_i h_i}} \sum_{\{\mathbf{k}_i\in \mathbb{N}^d\}| \sum_i\mathbf{k}_i=\mathbf{k}}\oint \Bigl(\prod_{j\ge 1}\frac{d\zeta_j}{2i\pi}\Bigr)\Bigl( \prod_{k\neq 1}\zeta_k^{q_k}\Bigr)\frac{\zeta_1^{q_1-1}}{x-\zeta_1}\prod_{J_i, \mathbf{k}_i}W_{\vec{J_i}}^{h_i}(\zeta_{\vec{J_i}}, x_{\mathbf{k}_i}) =0.
\end{align} 
The generating functions are now indexed by $d$-uplets living in $\mathbb{N}^d$ and this reflects the fact that there are several matrices. We indeed have,
\begin{eqnarray}
&W_{\mathbf{k}}(\underbrace{x^{(1)}_1,\cdots, x^{(1)}_{k_1}}_{k_1 \mbox{ times}}, \cdots , x^{(d)}_1, \cdots, x^{(d)}_{k_d}) \nonumber \\&= \sum_{\mbox{all }p^{(i)}_j\ge 0}\frac{\langle \tr(M_1^{p^{(1)}_1})\cdots \tr(M_1^{p^{(1)}_{k_1}})\tr(M_2^{p_1^{(2)}}) \cdots \tr(M_2^{p_{k_2}^{(2)}}) \cdots \tr(M_d^{p^{(d)}_1})\cdots \tr(M_d^{p^{(d)}_{k_d}})\rangle_c}{x^{(1)}_1{}^{p^{(1)}_{1}+1}\cdots x^{(1)}_1{}^{p^{(1)}_{k_1}+1}\cdots x^{(d)}_1{}^{p_1^{(d)}+1} \cdots x^{(d)}_{k_d}{}^{p^{(d)}_{k_d}+1}}.
\end{eqnarray}
These new $W$'s expand topologically in $1/N$ thanks to the special properties of this model in dimension six. This is how we obtain the $W_{\mathbf{k}}^g$ appearing in \eqref{eq:LEgeneral}. We indeed have $W_{\mathbf{k}}=\sum_{g\ge 0}N^{2-2g-|\mathbf{k}|}W^g_{\mathbf{k}}$. They are generating functions of colored stuffed maps of genus $g$, 'stuffed' meaning that cells of any topology are allowed\footnote{To be more precise, here the number of boundaries of the cells is bounded by $d$ the dimension of the original tensor model.}. 

\subsection{Spectral curve and Givental decomposition}

These loop equations also encode the geometry of an affine plane curve. However in this case the described surface is not a sphere but rather a collection of discs with two marked points on each of them. In fact, in \cite{Dartois}, it has been shown that the tensor model can be described in a Givental fashion, using the action of a differential operator $\hat{\mathcal{O}}$ on a product of Hermitian one matrix models
\begin{equation}
\mathcal{Z}[N,\lambda] =\exp(\hat{\mathcal{O}})\prod_{i=1}^d Z_{1MM}[\{t_p\}].
\end{equation}
This result, if derived only for $\mathcal{Z}$ in \cite{Dartois}, also extends to the case of $Z$. Moreover the resulting model is Gaussian at leading order, so that the resulting spectral curve is made of a union of Gaussian spectral curves, turning it into a disconnected spectral curve whose connected components are the spectral curves of  Gaussian models. Here, it is a union of discs with two marked points. 
The generalized Givental decomposition thus amount to describe the model in terms of quantities that are defined on each connected component of the spectral curve, but not necessarily on the full spectral curve. Each connected component is indeed described by the equation for the leading order $1$-point resolvent of the matrix of a given color $i$. Taking into account all the colors leads to the full disconnected spectral curve. However, one points out that if disconnected, the components are not independent one from another. As we indeed noticed sooner the choice of $W_1^0$ for each color corresponds to a choice of complex structure for the spectral curve, but in the tensor model case, the equation for $W_{e_i}^0$ involves constant terms that are fixed by the choices made for the $W_{e_j}^0$ for $j\neq i$:
\begin{equation} 
W^0_{e_i}(x)^2-(1-\alpha^2)xW^0_{e_i}(x)+\alpha^2 W_{e_i}^0(x)\sum_{j\neq i}\oint\frac{d\zeta_j}{2i\pi}\zeta_jW_{e_j}^0(\zeta_j)+(1-\alpha^2)=0.
\end{equation}
Combinatorial arguments on the Feynman graphs of the model lead to impose that the integral term is zero in this case. However, this argument comes from the fact that we consider the model as a formal integral, and could be false in general.   
If these terms do not change the topology of the curve, they potentially change the complex structure, for instance the position of the marked points can be changed.       
In the end the spectral curve associated to the model in dimension $d=4k+2$ for $k\in \mathbb{N}^*$ is described by the following data
\begin{proposition}
The spectral curve of the resummed quartic melonic tensor model is given by $S=(\Sigma, \omega_1^0=ydx,\omega_{2}^0 )$, where
\begin{itemize}
\item $\Sigma$ is a collection of $d$ discs with two marked points $\Sigma=\bigcup_{i=1}^d \mathbb{D}_i\{\pm 1\}$. Using the same uniformizing coordinates for each discs these marked points are located at $\pm 1$.
\item $\omega_1^0=\sum_i\Theta(z\in\mathbb{D}_i )\frac{(z-1/z)}{z^2}dz =\frac{(z-1/z)}{z^2}dz$ with $y=\sqrt{1-\alpha^2}\frac{dz}{z}$ and $dx=\frac{(z-1/z)}{\sqrt{1-\alpha^2} z}dz$.
\item $\omega_2^0(z_1,z_2)=\frac{\delta_{A,B}dz_1\otimes dz_2}{(z_1-z_2)^2}-\frac{\alpha^2(d-1)}{d(2\alpha^2-\alpha^4)+\alpha^4-\alpha^2-1} \frac{dz_1\otimes dz_2}{z_1^2z_2^2}$ for $z_1$ in the connected component $A$ while $z_2$ is in a connected component $B$ of $\Sigma$. 
\end{itemize}
\end{proposition}
To make the connection with the notations used before we simply notice that it is sufficient to add an index to each variables $z$ to specify in which connected disc they live, this additional index is their color. Doing so we recover that 
\begin{align}
&\omega^0_{2e_i}(z_1^{(i)},z_2^{(i)})=\frac{dz_1^{(i)}\otimes dz_2^{(i)}}{(z_1^{(i)}-z_2^{(i)})^2}-\frac{\alpha^2(d-1)}{d(2\alpha^2-\alpha^4)+\alpha^4-\alpha^2-1} \frac{dz_1^{(i)}\otimes dz_2^{(i)}}{(z_1^{(i)})^2(z_2^{(i)})^2} \\
&\omega^0_{e_i+e_j}(z_1^{(i)},z_1^{(j)})=-\frac{\alpha^2(d-1)}{d(2\alpha^2-\alpha^4)+\alpha^4-\alpha^2-1} \frac{dz_1^{(i)}\otimes dz_1^{(j)}}{(z_1^{(i)})^2(z_1^{(j)})^2}.
\end{align}
\section{A Blobbed Topological Recursion for The QMTM}
One shows that the $\omega_{\mathbf{k}}^g$ of the $d=4k+2$ dimensional QMTM satisfy a version of the topological recursion that is closely related to the one introduced by Gaetan Borot and Sergei Shadrin in \cite{BorotBlobbed}. These are results that are not published yet, even though they are proved and will be published in \cite{BonDar}. From any $\omega_{\mathbf{k}}^g$ we define two objects by the following relations
\begin{align}
&\omega_{\mathbf{k}}^g=\iPop \omega_{\mathbf{k}}^g +{}^iH \omega_{\mathbf{k}}^g\\
&\iPop \omega_{\mathbf{k}}^g=\sum_{\pm}\underset{z\rightarrow\pm 1 }{\textrm{Res}}{}^iG(z,z_1^{(i)})\omega_{\mathbf{k}}^g.
\end{align}
Here we denote $i$ the color of the first variable of $\omega_{\mathbf{k}}^g$. In the r.h.s. of the second line $\omega_{\mathbf{k}}^g$ depends on $z$ instead of $z_1^{(i)}$ (both $z$ and $z_1^{(i)}$ live in the same component). We also have that,
\begin{equation}
{}^iG(z,z_0)=-\int^z\omega_{2ei}^0(\cdot,z_0)
\end{equation}
for both $z$ and $z_0$ living on the $i^{th}$ discs. This quantity $\iPop \omega_{\mathbf{k}}^g$ satisfies a recursive formula.
\begin{proposition}
We have 
\begin{equation}
\iPop_{z_0}\omega_{e_i+\mathbf{k}}^g(z_0,z_{\mathbf{k}})=\sum_{\pm} \underset{z\rightarrow \pm 1}{\mbox{Res}} {}^iK(z, z_0) \tilde{Q}^g_{e_i+\mathbf{k}}(z;z_{\mathbf{k}})
\end{equation}
for some $\mathbf{k}$, $z_0$ living in the $i^{th}$ discs. $\tilde{Q}^g_{e_i+\mathbf{k}}(z_0;z_{\mathbf{k}})$  is the consistent generalization of the $\tilde{\mathcal{Q}}^g_n(z_1;z_2,\ldots, z_n) $. ${}^iK(z, z_0)$ is the kernel locally defined on each connected components by the use of $\omega_{e_i}^0$ and $\omega_{2e_i}^0$. 
\end{proposition}
In order to give a recursive formula for the whole $\omega_{\mathbf{k}}^g$ one needs to introduce possibly unknown quantities, called \emph{initial conditions}. They are differential forms denoted $\phi_{\mathbf{k}}^g
$, whose indices have the same role than the indices of $\omega_{\mathbf{k}}^g$. If one is able to provide an expression for these quantities one is, at least in principle, able to give an explicit expression for all $\omega_{\mathbf{k}}^g$. The analogue quantities introduced in \cite{BorotBlobbed} can, in principle, be computed in the case of the multi-trace Hermitian one matrix model described by \cite{BoStuffed}. In our case an efficient representation is still missing, the difficulty really comes from the fact that there are several connected components in the spectral curve. However we present a formula that applies to our situation, assuming the $\phi_{\mathbf{k}}^g$ are efficiently generated. \\

\bigskip

If one defines \cite{BonDar}
\begin{definition}
A normalized solution of loop equations $\omega_{ne_i}^{g,0}$ with $2g-2+n>0$ satisfies by definition
\begin{equation}
\omega_{ne_i}^{g,0}(z_0, I)=\sum_{\pm 1}\underset{z\pm 1}{Res}{}^iK(z,z_0)\Bigl[ \omega_{(n+1)e_i}^{g-1,0}(z,\iota(z), I)+\sum_{\substack{J\subseteq I \\ 0\le h \le g }}^{'}\omega_{(|J|+1)e_i}^{g,0}(z,J)\otimes\omega_{(|I-J|+1)e_i}^{g,0}(\iota(z),I-J)\Bigr].
\end{equation}
\end{definition}
We can give a graphical description of the solution in terms of the \emph{normalized} forms and the $\phi_{\mathbf{k}}^g$. First consider the set of labelled graphs.
\begin{definition}
We define $\mathfrak{G}^g_{\mathbf{k}}(A,B)$ the set of graphs satisfying the following properties:
\begin{itemize}
 \item \emph{Vertices:} Graphs $\Gamma \in \mathfrak{G}^g_{\mathbf{k}}(A,B)$ are made of vertices $v$ that are of two types, either $\omega^{0}$ or $\phi$. They carry an integer label $h(v)$ called its genus such that the valency $d(v)$ of $v$ satisfies $2h(v)-2+d(v)>0$. Furthermore, vertices of type $\omega^0$ are labeled by an integer $c\in [\![1,d]\!]$ called its color while each $\phi$ vertex $v$ is labeled by a multiplet $\mathbf{q} \in \mathbb{N}^{d}$, its colors, such that the valency of $v$ is $d(v)=|q(v)|$. Finally the component $q_i$ of $\mathbf{q}$ encodes the number of half edges of color $i$ incident to the vertex (\textit{i.e.} monocolored edges of color $i$ or bicolored edges of color $(j,i)$ such that the $i$ end is attached to the vertex).
 \item \emph{Edges are of two types:}
 \begin{enumerate}
  \item Monocolored edges: They are labeled by one integer $i\in [\![1,d]\!]$ called its color. Monocolored edges can only connect $\omega^0$ vertices to $\phi$ vertices.
  \item Bicolored edges: They are labeled by two integers $(i,j) \in [\![1,d]\!]^2 $, $i\neq j$ also called colors. These labels are to be seen as attached to the end of the edge. Bicolored edges of color $(i,j)$ can only connect $\omega^0$ vertices of color $i$ to $\omega^0$ vertices of color $j$.   
  \end{enumerate} 
  \item \emph{Leaves (unbounded edges):} There are $|\mathbf{k}|$ unbounded edges/leaves. Denote $\mathcal{L}$ the set of leaves. It can be written $\mathcal{L}=\bigsqcup_{i=1}^d \mathcal{L}_i$ with $\mathcal{L}_i=[\![1,k_i]\!]$ being the set of leaves of color $i$. Moreover $\mathcal{L}_i$ splits into the set of leaves with label $a$ and the one with label $b$ \textit{i.e.} $\mathcal{L}_i= A_i \sqcup B_i$. Then,
  \begin{enumerate}
 \item Each leaf of color $i$ with label $a$ can be of two types: 
  \begin{description}
   \item[a.] Either a monocolored leaf. In this case, it is incident to a $\phi_{\mathbf{q}}$ vertex with $q_i\ge 1$.
   \item[b.] Or a bicolored leaf. In this case, it is incident to a $\omega^0$ vertex of color $l$ such that the bicolored leaf has bicoloration $(l,i)$ and the $l$ end is attached to the $\omega^0$ vertex while the $i$ end is left free. 
  \end{description}
  \item Each leaf of color $i$ with label $b$ is incident to a $\omega^0$ vertex of color $i$
  \end{enumerate}
  \item $\omega^0$ vertex have at least a leaf. Moreover each $\omega^0$ vertex must have at least one monocolored leaf.
  \item $\Gamma$ is connected and $b_1(\Gamma)+\sum_v h(v)=g$. 
\end{itemize}
\end{definition}
To each of these graphs is attached a weight in such a way that we can compute
\begin{theorem}\label{thm:multicolorBgraphs}
Given $A=(A_1,A_2,\ldots,A_d)$ and $B=(B_1,B_2,\ldots,B_d)$ two multiplets of subsets of the sets $[\![1,k_i]\!]$ such that $A_i\sqcup B_i = [\![1,k_i]\!]$, we have:
\begin{equation}
H_A \ P_B \ \omega_{\mathbf{k}}^g=\sum_{\Gamma\in \mathfrak{G}^g_{\mathbf{k}}(A,B) }\frac{\varpi_{\Gamma}(z_{\mathbf{k}})}{|\textrm{Aut}(\Gamma)|},
\end{equation}
where $\varpi(z_{\mathbf{k}})$ is a weight associated to the graph $\Gamma$.
\end{theorem}
The weight $\varpi(z_{\mathbf{k}})$ is computed in the following way. Set $\Gamma$ a graph in $\mathfrak{G}^g_{\mathbf{k}}(A,B)$. The weight of $\Gamma$ is made out of a product of local weight to which we apply a certain pairing. Each vertex represent a certain local weight. This weight is of type $\omega^{h(v),0}_{d(v)e_i}(\{z_e^i\})$ for a vertex $v$ of color $i$ of type $\omega^0$ with valency $d(v)$. On the other hand this local weight is given by $\phi_{\mathbf{q}(v)}^g$ for a vertex of type $\phi$ while $\mathbf{q}(v)$ is determined by the colors of the half edges incident to it. Now split the set of edges of $\Gamma$ into the set of colored edges and the one of bicolored edges.  The same shall be done for the set of leaves.

\medskip

First consider the set of edges. Attach a variables $z_e^i$ to each edge $e$ of color $i$, while attaching a duo of variables $(z_e^i,z_e^j)$ to each bicolored edge of color $(i,j)$. For each edge of color $i$ incident to both a vertex $v$ of type $\omega^0$ of color $i$ and a $\phi$ vertex $v'$ with $q_1\ge 1$ we compute the following pairing:
\begin{equation}
\bigl\langle\ \omega^{h(v),0}_{d(v)e_i}(\ldots,z_e^i)\phi_{\mathbf{q}(v')}^{h(v')}(\ldots,z_e^i,\ldots) \bigr\rangle_{z_e^i}=\sum_{\pm} \underset{z_e^i \rightarrow \pm 1}{\textrm{Res}} \omega^{h(v),0}_{d(v)e_i}(\ldots,z_e^i) \int_{\pm 1}^{z_e^i}\phi_{\mathbf{q}(v')}^{h(v')}(\ldots,z_e^i,\ldots).
\end{equation} 
Consider the case of bicolored edges. A bicolored edge labeled $e$ of color $(i,j)$ is attached to a $\omega^0$ vertex $v$ of color $i$ at one of its end and to a $\omega^0$ vertex $v'$ at its other end. Both end $i,j$ of the edge carry a variable $z_e^i$ and $z_e^j$. The corresponding pairing function is:
\begin{equation}
\bigl\langle\ \omega^{h(v),0}_{d(v)e_i}(\ldots,z_e^i)\omega^{h(v'),0}_{d(v')e_j}(z_e^j,\ldots) \bigr\rangle_{z_e^i,z_e^j}=\sum_{\pm} \underset{z_e^i\rightarrow \pm 1}{\textrm{Res}}\underset{z_e^j\rightarrow \pm 1}{\textrm{Res}}\omega^{h(v),0}_{d(v)e_i}(\ldots,z_e^i)\omega^{h(v'),0}_{d(v')e_j}(z_e^j,\ldots)\int_{\pm 1}^{z_e^i} \int_{\pm 1}^{z_e^j}\omega^0_{e_i+e_j}(\cdot,\cdot).
\end{equation} 

\medskip

Consider the leaves of the graph. Monocolored leaves are incident to $\omega^0$ vertices. They represent variables appearing in the weight $\varpi_{\Gamma}(z_{\mathbf{k}})$. They have the color of the variables they represent. However, bicolored leaves are associated to another pairing. They also represent some variable $z^{(j)}$ attached to their free end with color the one labeling this free end ($j$). Moreover, they are incident to a $\omega^0$ vertex $v$ of some color $i$, this is associated to the following pairing,
\begin{equation}
\bigl\langle\omega_{d(v)e_i}^{h(v), 0}(\ldots, z_e^i)\omega_{e_i+e_j}(z_e^i,z^{(j)})\bigr\rangle_{z_e^i}=\sum_{\pm 1}\underset{z_e^i\rightarrow \pm 1}{\textrm{Res}}\omega_{d(v)e_i}^{h(v), 0}(\ldots, z_e^i)\int_{\pm 1}^{z_e^i}\omega_{e_i+e_j}(z_e^i,z^{(j)}).
\end{equation}
From Theorem \ref{thm:multicolorBgraphs}, it is possible to compute ${}^iH \omega_{\mathbf{k}}^g$ for any $(\mathbf{k}, g)$ assuming that one knows about the normalized $\omega_{n'e_i}^{g',0}$ for any $(g',n')$ and the $\phi_{\mathbf{k}'}^{g'}$ satisfying respectively $2g'-2+n'<2g-2+|\mathbf{k}|$ and $2g'-2+|\mathbf{k}'|<2g-2+|\mathbf{k}|$. This is indeed the case for the normalized forms that are recursively computed from a topological recursion which just needs as an input $\omega_{e_i}^0, \omega_{2e_i}^0$. But this is not true in general that we can generate the $\phi_{\mathbf{k}'}^{g'}$.
In some sense the idea we described can be seen as either a generalization or a specialization (depending on your tastes) of the formalism introduced in \cite{BoStuffed}. In the formalism introduced in \cite{BoStuffed}, the initial conditions are named this way because this is something one has to add as an input to the Blobbed Topological Recursion to obtain the solutions of the mathematical problem one wants to solve. But depending on this specific problem, it can or cannot be possible to write these initial conditions in a closed way. The current question is, is it possible to do so in the context of the QMTM?

\bigskip

As we saw in this part the generalization of the topological recursion ideas to the simplest tensor models already involves using/generalizing state of the art techniques for the topological recursion. In spite of these efforts, the colored version of the $W$'s computed here are generating functions for a very restricted set of observables of the original tensor models (melonic cyclic graphs). It is not known at the moment how to obtain a more general set of observables. Moreover, works need to be done in dimension $d\neq 4k+2$. Of course, it is possible to force the $\omega_{\mathbf{k}}^g$ to depend on $N$. This would circumvent the problem of the non-topological nature of the $1/N$ expansion at the level of the analytical properties of the $\omega_{\mathbf{k}}^g$. This solution has the major disadvantages of spoiling the ordering provided by the $1/N$ expansion and being conceptually not satisfactory. The question is whether or not there exists a more natural setting that would extend the topological recursion idea and naturally contain the non-topological case (maybe simplifying when the topological situation is recovered)?
\section{Why is it difficult to compute the $\phi_{\mathbf{k}}^g$?} 
In the $1$-cut case for multi-trace Hermitian matrix model, one takes advantage of the fact that the corresponding linear loop equations for $2g-2+n>0$ writes 
\begin{equation}
(S+O)\omega^g_n=d_{z_1}V_n^g(z_1; z_2,\ldots, z_n),
\end{equation}        
where $V_n^g(z_1; z_2,\ldots, z_n)$ is a holomorphic function of $z_1$ around the marked points of the spectral curve whose expression is known and only depends on $\omega_{n'}^{g'}$ with $2g'-2+n'<2g-2+n$ and $O$ is a linear operator constructed out of $W_1^0$ and the planar $2$-point potential. Moreover the corresponding bi-differential satisfies
\begin{equation}
(S+O)\omega^0_2=\frac{dx_1\otimes dx_2}{(x(z_1)-x(z_2))^2}.
\end{equation}  
From this very fact we have that $(S+O)P\omega_g^n=0$ which implies $(S+O)H\omega^g_n=d_{z_1}V_n^g(z_1; z_2,\ldots, z_n)$. It is possible to construct a second, simple, expression for $H\omega^g_n$ by just convoluting $V_n^g(z_1; z_2,\ldots, z_n)$ with $\omega_2^0$. Then by acting with the $H$ operator on all remaining variables $z_2,\ldots, z_n$ of this expression, one gets, a quite involved, expression for the $\phi_{n}^g$ that only depends on objects with greater Euler characteristic, so that they can in principles be computed at any order \cite{BoStuffed,BorotBlobbed}.

\bigskip 

In the case of tensor models, the situation changes a bit, and prevents this technique to be applied efficiently. Indeed, the linear loop equations have in general the form
\begin{equation}
S_{z_O}\omega_{\mathbf{k}}^g+O\sum_{\substack{\mathbf{k}_i\\ |\mathbf{k_i}|=|\mathbf{k}|}}c_i\omega_{\mathbf{k}_i}^g=d_{z_0}V_{\mathbf{k}}^g(z_0; \ldots).
\end{equation} 
This linear loop equation also involves the action of an operator that looks like the $O$ of the one matrix case, but, this operator is now acting on a linear combination\footnote{some coefficients $c_i$ may vanish.} of $\omega_{\mathbf{k}_i}^g$ that depends on $\mathbf{k}$. If in some cases, it is possible to compute the term $O\sum_{\substack{\mathbf{k}_i\\ |\mathbf{k_i}|=|\mathbf{k}|}}c_i\omega_{\mathbf{k}_i}^g$ in term of $\omega_{\mathbf{k}}^g$ this new expression add new terms in the equation that depends on the Taylor expansion of the $\omega_{\mathbf{k}'}^{g'}$ with $2g'-2+|\mathbf{k}'|<2g-2+|\mathbf{k}|$. Moreover obtaining this new expression involves coming back to the original loop equations and study their expansion at $x_0=\infty$. From this expansion we can extract the data we need to compute $O\sum_{\substack{\mathbf{k}_i\\ |\mathbf{k_i}|=|\mathbf{k}|}}c_i\omega_{\mathbf{k}_i}^g$ in term of a new linear operator $O_{\mathbf{k}}^g$ acting on $\omega_{\mathbf{k}}^g$ and a change of the function $V_{\mathbf{k}}^g$. The problem of this method is that it is not very efficient, and the relative computability of the low order cases relies heavily on the fact that we focused on the symmetric QMTM (\textit{i.e.} all coupling constant of the interactions terms are the same). 

\bigskip

What does it means for the model? The current state of the art is that the model in dimensions $d=4k+2$ indeed satisfies a version of the blobbed topological recursion. However it is only an abstract tool as it is not yet possible to construct the solution efficiently just from the graphical expansion given by this version of the blobbed topological recursion. If one forgets about the efficiency (as compared to the blobbed topological recursion of the $1$-cut multi-trace Hermitian one matrix model\footnote{which is already very involve when compares to the usual topological recursion applied to the $1$-cut Hermitian one matrix model.}), one can obtain, in principle the expression of the $\phi_{\mathbf{k}}^g$ as their calculation always involves objects that are already known from lower order computations. However, there is no automatic method to describe them yet. In general, one needs to invert an affine operator at each order, and the size of this operator grows with the order of the $\phi$ one wants to compute. 

\bigskip

However, the good news is that this colored version of the blobbed topological recursion can be understood as the blobbed topological recursion applied to the case of a multi-cut disconnected spectral curve naturally generated by the tensor model case. Firstly, it is the first example of a natural model generating such a curve. Secondly, it implies that all the structural properties of the blobbed topological recursion are satisfied. In particular, it implies that the subset of cyclic melonic observables are related to combination of intersection numbers, so that the QMTM decomposes on Kontsevitch matrix integrals \cite{Kont1}, which is a second form of generalized Givental decomposition. These results allow tensor models to make connection with one of the most active part of mathematical physics nowadays.   

\section{Conclusion}

In this review we explained recent ideas that need to be developed in the context of tensor models. The Virasoro constraints in some matrix models allow one to reformulate the computation of some (generating functions of) observables of matrix models in terms of computations of correlation functions of conformal field theory (CFT). Can we extend this idea using the Lie algebra of constraints of tensor models?  Is there an analogue of CFT that carries this constraint algebra of tensor models as its algebra of symmetry? Is it linked to $\mathcal{W}$ algebras?

\bigskip

Matrix models associate to each piecewise linear collection of $\mathbb{S}^1$ an amplitude that is computed from combinatorial weights associated to each $PL$ cobordisms between these $\mathbb{S}^1$. From this point of view they are an incarnation of two dimensional Topological Field Theory. \\
Tensor models do the same but with more dimensions. For instance in the $d=3$ dimensional case, they associate to each collection of $PL$ surfaces an amplitude that is computed from combinatorial weight associated to each $PL$ cobordisms between them. But, from the liberty we have in choosing the $q_{B_{\bullet}}$, we can now also distinguish and weight the sub-manifolds of these surfaces. This let us think (without rigorous arguments) that three dimensional tensor models have the flavor of a categorification of matrix models, indeed the one dimensional sub-manifolds of the observables can be seen at the objects, the observables as the morphisms and the cobordims between observables as the $2$-morphisms. If one wants to generalize the idea of topological recursion to higher dimensional models such as tensor models, it would certainly be interesting to understand the Topological Recursion in the categorical language and then try to categorify it, while using tensor models as a guiding example for this categorification.

\bigskip

A more approachable series of problems is probably to extend the techniques developed in the case of the QMTM, and this independently of the possibility to find an efficient representation of the $\phi_{\mathbf{k}}^g$ or not. First of all, one should try to reformulate these results using the tensor variables. It could also be interesting to focus on the cases of larger family of melonic interactions. One could also try to find a specification of the $q_{B_{\bullet}}$ that leads to interesting properties for the thus defined generating functions.

From the purely topological recursion point of view, results concerning the fate of integrable structures in the blobbed framework would be interesting. If they certainly do not survive as they are, it may well be possible that they are deformed or generalized in some way. Such results could shed some light on the 'integrability/computability/solvability' properties of the QMTM (while being interesting on its own).            
 
\bigskip

The topological recursion has only recently been generalized to a class of tensor models\cite{BonDar}. This starts a new program whose goal is the study of topological recursion within the framework of tensor models. This program could help us further understand the mathematics underlying tensor models, while maybe revealing limits to the application of the topological recursion. If it is probably a very difficult program, it is possible to make small steps. One shall start by focusing on simple models, and by getting inspiration from what have been done in the context of matrix models and the topological recursion, for which there already is a rich mathematical apparatus, ranging from very practical mathematical tools (that are well suited to computations) from very abstract ideas (that are more suited to understanding the generalized structures). 

\nocite{*}
\bibliographystyle{alpha}
\bibliography{biblio}

\end{document}